\documentstyle[epsf,sprocl]{article}

\input{psfig}
\def\be{\begin{equation}}
\def\ee{\end{equation}}
\def\bea{\begin{eqnarray}}
\def\eea{\end{eqnarray}}

\begin{document}

\title{ON THE STRUCTURE AND NATURE OF DARK MATTER HALOS}

\author{ ANDREAS BURKERT}

\address{Max-Planck-Institut f\"ur Astronomie, K\"onigstuhl 17,\\
69117 Heidelberg, GERMANY}

\author{ JOSEPH SILK}

\address{Department of Astrophysics, NAPL, Keble Rd, Oxford OX1 3RH, UK,
and Departments of Astronomy and Physics, University of California, \\
Berkeley, CA 94720, USA}

\maketitle\abstracts{The structure of dark matter halos as predicted from
cosmological models is discussed and compared with observed rotation curves
of dark matter-dominated dwarf galaxies. The theoretical models predict that
dark matter halos represent a one-parameter family with a universal density
profile.  Observations of dark matter-dominated rotation curves indeed
confirm the universal structure of dark halos. They are even in excellent
agreement with the expected mass-radius scaling relations for the currently
favoured cosmological model (standard cold dark matter with $\Omega_0=0.25$
and $\Omega_{\Lambda}=0.75$). The rotation profiles  however disagree
with the predicted dark matter density distributions.  Secular processes
which might affect the inner halo structure do not seem to provide a good
solution to this problem.  We discuss, as an alternative, the possibility
that dark halos consist of two separate components, a dark baryonic and a
dark non-baryonic component.}

\section{Introduction}

Cosmological  models of hierarchical merging in a cold dark matter universe
are in some difficulty. High-resolution N-body simulations (Navarro et al. 1996a; NFW)
have shown that the density profiles $\rho_{NFW}$ of virialized dark matter halos should
have a universal shape of the form

\begin{equation}
\rho_{NFW} (r) = \frac{3 H_0^2}{8 \pi G} \frac{\delta_c }{(r/r_s)(1+r/r_s)^2}
\end{equation}

\noindent where $r_s$ is a characteristic length scale
and $\delta_c$ a characteristic
density enhancement. The two free parameters $\delta_c$ and $r_s$
can be determined from the halo concentration c and the virial mass $M_{200}$

\begin{eqnarray}
\delta_c & = & \frac{200}{3} \frac{c^3}{\ln (1+c) - c/(1+c)}  \\
r_s      & = & \frac{R_{200}}{c} = \frac{1.63 \times 10^{-2}}{c}
\left(\frac{M_{200}}{M_{\odot}} \right)^{1/3} h^{-2/3} kpc.
\end{eqnarray}

\noindent where $R_{200}$ is the virial radius, that is the radius inside
which the average overdensity is 200 times the critical density of the universe
and $M_{200}$ is the mass within $R_{200}$.

For any  particular cosmology there also exists  a  good correlation
between c and $M_{200}$
which results from the fact that dark halo densities reflect the density
of the universe at the epoch of their formation (NFW,
Salvador-Sol\'{e} et al. 1998) and that halos of
a given mass are preferentially assembled over a narrow range of redshifts.
As lower mass halos form earlier, at times when
the universe was significantly denser, they are more centrally concentrated.
NFW have published concentrations for dark matter
halos in the mass range of $3 \times 10^{11} M_{\odot} \leq M_{200} \leq 3 \times
10^{15} M_{\odot}$ which can be well fitted by the following power-law functions:

\begin{eqnarray}
c & = & 8.91 \times 10^2 \left(\frac{M_{200}}{M_{\odot}} \right)^{-0.14}
\ \ \ \ \ {\it for} \ \ \  {\rm SCDM}  \\
c & = & 1.86 \times 10^2 \left(\frac{M_{200}}{M_{\odot}} \right)^{-0.10}
\ \ \ \ \ {\it for} \ \ \ {\rm CDM}\Lambda \nonumber
\end{eqnarray}

\noindent where SCDM denotes  a standard biased cold dark matter model
with $\Omega_0$=1, h=0.5, $\sigma_8$=0.65 and CDM$\Lambda$ denotes a low-density universe
with a flat geometry and a non-zero cosmological constant, defined by
$\Omega_0$=0.25, $\Omega_\Lambda$ = 0.75, h=0.75, $\sigma_8$=1.3.
Note that the universal profile (equation 1) and the scaling relations (equation 4)
have only been determined from simulations for  halo masses as small as
$M_{200}=10^{11}M_{\odot}$, but there is no reason to believe that these results
would not be valid for halos which are, say,  one order of magnitude lower in mass.
In summary, dark matter halos represent a one-parameter family, with
their density distribution being determined completely by their virial mass $M_{200}$.

The universal character of dark matter profiles and the validity of
the NFW-profile for different cosmogonies has  been verified
by many N-body calculations (e.g. Navarro et al. 1997, 
Cole \& Lacey 1997, Tormen et al. 1997, Tissera \& Dominguez-Tenreiro 1998, Jing 1999).
In one of the highest resolution simulations to date,
Moore et al. (1998, see also Fukushige \& Makino 1997) found good agreement with the
NFW profile (equation 1) at and outside of $r_s$. Their simulations did however lead to
a steeper innermost slope $\rho \sim r^{-1.4}$ which extends all the way
down to their resolution limit of 0.01 $r_s$.

On the analytical side, early spherically symmetric collapse models by Gunn \& Gott (1972) studied the collapse of a uniformly overdense region. Gott (1975) and Gunn (1977)
investigated secondary infall onto already collapsed density perturbations and
predicted $r^{-9/4}$ profiles.
Fillmore \& Goldreich (1984)  found self-similarity solutions for
the secondary infall models.
Hoffman \& Shaham (1985) took into account more realistic Gaussian initial conditions
and predicted sharp central density peaks of the form $\rho \sim r^{-2}$.
An updated version of these models by Krull (1999) abandoned self-similarity
and 
explicitly took
into account the hierarchical formation history. His models lead to
excellent agreement with the NFW-profile in the radius range $0.5 r_s \leq r \leq 10 r_s$.

\begin{figure}[h]
\centerline{\psfig{figure=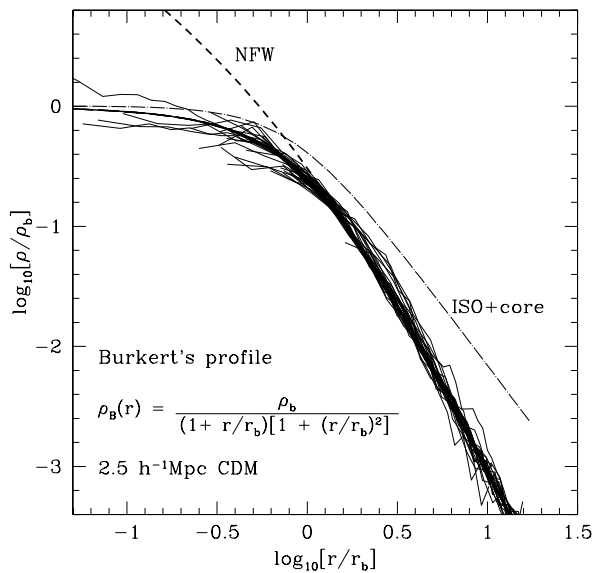,height=8cm}}
\caption{The density distributions of dark  halos in a cold dark matter simulation of KKBP (thin solid
lines) are compared with the Burkert profile (thick solid line) that provides a good
fit to the observed rotation curves of dark matter-dominated dwarf galaxies. The thick
dashed line shows the NFW profile which predicts too much dark matter mass inside 
$r_b$. The dot-dashed curve shows an isothermal profile with a finite density core which 
fails in the outer regions where it decreases as $r^{-2}$. }
\end{figure}

In a different series of very high-resolution models, using a new adaptive
refinement tree N-body code,
Kravtsov et al. (1998, KKBP) found significant deviations from the NFW profile or an
even steeper inner power-law density distribution for $r \leq 0.5 r_s$.
In this region their dark matter profiles show a  substantial scatter
around an average profile that is characterized by a
shallow central cusp with $\rho \sim r^{-0.3}$.
Although the scatter is large, this result is in clear contradiction to the simulations
of Moore et al. (1998) with equally high central resolution.
Figure 1 (adopted from Primack et al. 1998)  
compares the NFW-profile (dashed line) with the profiles of dark
matter halos of KKBP (thin solid lines).

\section{The dark matter halo of DDO 154}

DDO 154 (Carignan \& Freeman 1988) is one of the most gas-rich galaxies known with
a total H I mass of $2.5 \times 10^8 M_{\odot}$ and an inner stellar component of
only $5 \times 10^7 M_{\odot}$. Recently, Carignan \& Purton (1998) measured the
rotation curve of its extended H I disk all the way out to 21 optical disk scale
lengths. As the rotation curve, even in the innermost regions, is almost completely
dominated by dark matter, this galaxy provides an ideal laboratory for testing the
universal density profile predictions of cosmological models.

\begin{figure}[h]
\centerline{\psfig{figure=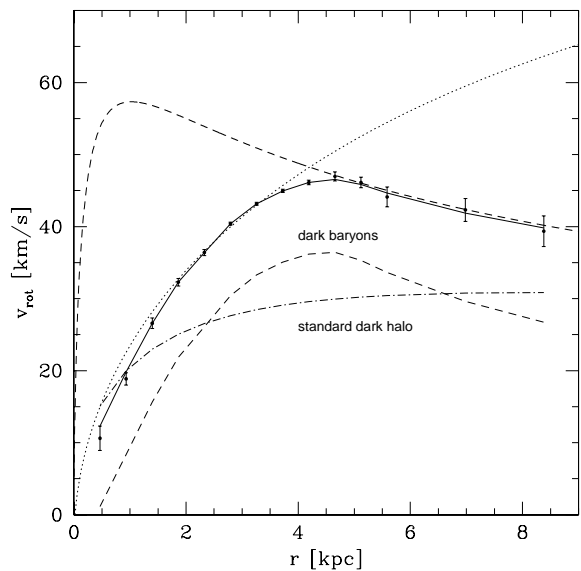,height=8cm}}
\caption{The dark matter rotation curve of DDO 154 with error bars.
The dotted line is a fit to the inner parts, adopting a NFW profile. The dashed
curve shows a fit to the outer regions. The solid line shows the fit which is
achieved with a two-component model where the lower dashed line shows the
contribution of the dark baryonic component and the lower dot-dashed curve
is the standard dark matter halo contribution adopting a NFW profile.}
\end{figure}

Figure 2 shows the dark matter rotation curve of DDO 154 and compares it with
the NFW profile. Note that the maximum rotational velocity $v_{max} \approx$ 47 km/s
is reached at a radius of $r_{max} \approx$ 4.5 kpc, beyond which the rotation
decreases again. Fitting the inner regions (dotted line) has been known to pose
a problem (Flores \& Primack 1994, Moore 1994, Burkert 1995). However an even larger
problem exists in the outermost regions where far too much dark matter would be expected.
The dashed line in figure 2 shows a fit to the outer regions. In this case, the dark
matter excess in the inner regions is unacceptably large. We conclude that
 the well-studied
dark matter rotation curve of DDO 154 is far from   agreement with NFW profiles.

We can also compare the observed location $r_{max}$ and the value 
$v_{max}$ of the observed maximum rotational velocity with predictions of a SCDM model.
Adopting a NFW profile, the virial radius is determined by $R_{200}$= 0.5 c r$_{max}$. The virial
mass is then given by the relation

\begin{equation}
1.63 \times 10^{-2} \left(\frac{M_{200}}{M_{\odot}} \right)^{1/3} h^{-2/3} =
\left(\frac{R_{200}}{\rm kpc} \right) = 0.5 c \left(\frac{r_{max}}{\rm kpc} \right) .
\end{equation}

Adopting the SCDM model with h=0.5 and inserting equation (4) for $c$ one obtains

\begin{equation}
M_{200}^{SCDM} = 9 \times 10^8 \left(\frac{r_{max}}{\rm kpc} \right)^{2.11} M_{\odot}
\end{equation}

For DDO 154 with $r_{max}$ = 4.5 kpc we find $M_{200} = 2.1 \times 10^{10} M_{\odot}$,
$R_{200}$ = 72 kpc and $c=31.9.$
 For these halo values, the predicted maximum rotational
velocity  would be

\begin{equation}
v_{max} = 0.465 \left( \frac{c}{ln(1+c)-c/(1+c)} \right)^{1/2} h
\left(\frac{R_{200}}{\rm kpc} \right) = 60 \  {\rm km/s}
\end{equation}

\noindent which is a factor 1.3 larger than observed.

Adopting instead the CDM$\Lambda$ model with h=0.75, a similar calculation leads to

\begin{equation}
M_{200}^{CDM \Lambda} = 3 \times 10^8 \left(\frac{r_{max}}{\rm kpc} \right)^{2.3} M_{\odot}
\end{equation}

\noindent and therefore to $M_{200} = 9.5 \times 10^9 M_{\odot}$,
$R_{200}$=42 kpc, c=18.7 and $v_{max}$=44.4 km/s which is in excellent in
agreement with the observations (47 km/s), especially if one notes that
equation 4 has been verified only for viral masses of $M_{200} > 10^{11}
M_{\odot}$.

In summary, the radius and mass scale of DDO 154 as determined from the value
and location of its maximum rotational velocity is in perfect agreement with
the predictions of the currently most favoured cosmological models
(CDM$\Lambda$). The inferred dark matter density distribution is however quite
different.

\section {The universality of observed dark matter mass profiles.}

DDO 154 is not a peculiar case. Burkert (1995) showed that the dark matter
rotation curves of four dwarf galaxies studied by Moore (1994) have the same
shape which can be well described by the density distribution 

\begin{equation}
\rho_B(r) = \frac{\rho_b}{(1+r/r_b)[1+(r/r_b)^2]}.
\end{equation}

KKBP extended this sample to ten dark matter-dominated dwarf irregular
galaxies and seven dark matter-dominated low surface brightness galaxies. As
shown in figure 3 {\it all} have the same shape, corresponding to a density
distribution given by equation (9) and in contradiction to equation (1).

\begin{figure}[h]
\centerline{\psfig{figure=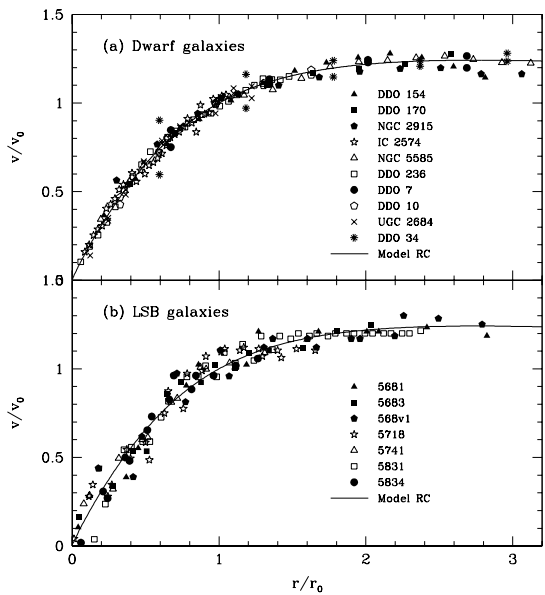,height=9cm}}
\caption{This figure, adopted from Primack et al. (1998) shows the dark matter rotation
curves of (a) ten dwarf irregular and (b) seven low surface brightness galaxies.
The solid line shows the profile proposed by KKBP which is nearly identical to $\rho_B$
(equation  9).}
\end{figure}

Equation 9 predicts a flat dark
matter core, the origin of which is difficult to understand in the context of hierarchical
merging (Syer \& White 1998) as lower-mass dark halos in general have high densities
and therefore spiral into the center of the more diffuse merger remnant, generating
high-density cusps.

There is  a fundamental difference in the kinematical properties of dark
matter halos described by NFW profiles and Burkert profiles (equation 9). 
Assuming isotropy and spherical symmetry in the inner regions, the Jeans equation 
predicts a velocity dispersion profile $\sigma (r)$ of NFW halos that decreases 
towards the center as $\sigma \sim r^{1/2}$ (see also the simulations of
Fukushige \& Makino 1997) whereas Burkert halos have isothermal cores with constant
velocity dispersion. Again, non-isothermal, kinematically cold dark matter 
cores might  be expected
in hierarchical merging scenarios as the denser clumps that sink towards the center
of merger remnants have on average smaller virial dispersions.

\section{On the origin of dark matter cores}

It is rather  unsatisfying  that numerical studies of dark matter
halos using different techniques lead to different results.  KKBP
find dark matter halo profiles that, on average,  can be well fitted by a Burkert
profile (Fig. 1) and therefore also provide a good fit to observed rotation curves.
The NFW-profiles, or even steeper inner density gradients (Moore et al. 1998)
seem to be in clear
conflict with observations (Fig. 2). Numerical resolution cannot be the answer
as the N-body simulations of Moore et al. (1998) have enough resolution to determine
the dark matter density distribution within 0.5 $r_s$, where any flattening should have 
been found. Instead, the authors find profiles that are even steeper than $r^{-1}$.
One should note that the results of KKBP have not yet been reproduced by other
groups, whereas dark profiles with central $r^{-1}$ profiles or even steeper cusps have
been found independently in many studies using different numerical techniques.
On the other hand, the high-resolution studies of KKBP sample
galactic halos, whereas most of the other studies simulate halos of galactic
clusters. Indeed, there exists observational evidence that the dark matter distribution
in clusters of galaxies is well described  by NFW profiles (Carlberg et al. 1997,
McLaughlin 1999). One conclusion therefore could be that dark matter halos are
not self-similar but that their core structure does depend on their virial mass.
Because one is sampling different parts of the primordial
CDM density fluctuation power spectrum, from which the initial conditions are
derived,
it  is possible that the initial conditions could influence the final result.
For example, low mass dark halos have  initial fluctuations which result in
virialized velocity dispersions that are nearly
independent of mean density, and so a hierarchy of substructure would be
expected to be nearly isothermal.
For more massive halos typical of normal galaxies,  the velocity dispersion
varies as $\rho^{\frac{n-1}{3(n+3)}}$ where $n$ is the effective power-law power
spectrum index $\delta \rho / \rho \propto M^{-\frac{n+3}{6}}$
and $n \approx -2 $ on galaxy scales but $n \approx -3$ for dwarfs.
Low-mass dark halos could then indeed have isothermal, constant density cores 
whereas high-mass dark halos should contain
non-isothermal, cold power-law cores. However, even in this case, 
one still has the problem 
that the simulations lead to a much greater dispersion of the inner radial profiles
than expected from observed rotation curves.
Additional effects might therefore be important that have not been taken into account
in dissipationless cosmological simulations.

\subsection{Secular dynamical processes}

Cold dark matter cores with steep density cusps are very fragile and can easily
be affected by mass infall and outflow. This has been shown, for example, by
Tissera \& Dom\'{i}nguez-Tenreiro (1998) who included gas infall in their cosmological models
and found even steeper
power-laws in the central regions than predicted by purely dissipationless merging
due to the adiabatic contraction of the dark component.
In order to generate flat cores through secular processes, Navarro et al. (1996b)
proposed a scenario where after a slow growth of a dense gaseous disk the gas
is suddenly ejected. The subsequent expansion and  violent relaxation  phase 
of the inner dark matter
regions leads to a flattening of the core. This model has been improved by
Gelato \& Sommer-Larsen (1999) who applied it to DDO 154 and found that
it is not easy to satisfactorily explain the observed rotation curve even
for extreme mass loss rates. In fact,
it is unlikely that DDO 154 lost any gas, given its large gas fraction.
In addition, secular dynamical processes due to mass outflow would predict
inner dark matter profiles that depend sensitively on the detailed physics of
mass ejection and therefore should again  show  a wide range of density distributions, 
and these are not observed.

\subsection{A second dark and probably baryonic component}

The rotation curves of the galaxies shown in figure 3 clearly cannot be
understood by including only the visible component. It may well be
that  some
non-negligible and as yet undetected fraction of the total baryonic mass
contributes to their dark component, in
addition to the non-baryonic standard cold dark component that is considered
in cosmological models.

In fact, there is ample room for such a dark baryonic component.
Primordial nucleosynthesis requires a baryonic density of $\Omega_b h^2 \approx
0.015 \pm 0.008$ (Kurki-Suonio et al. 1997, Copi et al. 1995), whereas
the observed baryonic density for stellar and gaseous disks lies in the range
of $\Omega_d \approx 0.004 \pm 0.002$ (Persic \& Salucci 1992). 
Moreover modelling of Lyman alpha clouds at $z\sim 2-4$ suggests that all of
the baryons expected from primordial nucleosynthesis
were present in diffuse form, to within the uncertainties,
which may amount to perhaps a factor of 2 (Weinberg et al. 1997).
Hence dark baryons are required at low redshift. These may be in the form of
hot gas that must be mostly outside of systems such as the Local Group and
rich galaxy clusters (Cen and Ostriker 1999).
But an equally plausible possibility is that the dark baryons are
responsible for a significant fraction of the mass
in galaxy halos, as is motivated by arguments involving disk rotation curves
and halo morphologies (cf Gerhard and Silk 1996; Pfenniger et al. 1994).

This second 
dark baryonic component could be diffuse $H_2$ within the disks or some spheroidal distribution
of massive compact baryonic objects (MACHOs), comparable to
 those that have been detected
via gravitational microlensing events towards the Large Magellanic Cloud (LMC).
It is difficult to reconcile the inferred typical MACHO lens mass of $\sim 0.5 \pm 0.3$
M$_{\odot}$, as derived from the first 2.3 years of data for 8.5 million stars in
the LMC (Alcock et al. 1996), with ordinary hydrogen-burning stars or old white dwarfs
(Bahcall et al. 1994, Hu et al. 1994, Carr 1994, Charlot \& Silk 1996).
Brown dwarfs, substellar objects below the hydrogen-burning limit of 0.08 $M_{\odot}$
would be ideal candidates. Indeed, halo models can be constructed, e.g. by
assuming a declining outer rotation curve, for which the most likely MACHO mass
is 0.1 M$_{\odot}$ or less ( Honma \& Kan-ya 1999) with a MACHO contribution to
the total dark mass of almost 100\%. Freese et al. (1999) have however shown by deep
star counts using HST that faint stars and massive brown dwarfs contribute no more than 1\% of
the expected total dark matter mass density of the Galaxy, ruling out such a
low-mass population. 

A simple explanation of the MACHO mass problem has been presented by
Zaritsky \& Lin (1997) and Zhao (1998), who argued that the  MACHOs reside in a previously
undetected  tidal stream, located somewhere in front of the LMC. In this case the microlensing
events would represent stellar objects in the outer regions of the LMC and
would not be associated with a dominant dark matter component of the Milky Way.
This solution is supported by the fact that all lensing events toward the LMC and
SMC with known distances (e.g. the binary lensing event 98 LMC-9, or 98-SMC-1)
appear to be a result of self-lensing within the Magellanic Clouds.
However the statistics are abysmal: only two SMC events have been reported, 
and there are approximately 20 LMC events in all, of which two
have known distances. Moreover one can only measure distances for binary
events, and these are very likely due to star-star lensing. Finally, the SMC
is known to be extended along the line of sight, thereby enhancing the
probability of self-lensing.

Thus, while evidence for a possible dark baryonic component in the
outer regions of galaxies is small, such a component could still be the solution 
to the dark matter core problem.
Burkert \& Silk (1997) showed that the observed rotation curve of DDO 154 could
be reconciled with standard cosmological theories if, in addition to a standard
dark matter halo with an NFW-profile (corrected for adiabatic contraction),
a separate and centrally condensed dark baryonic component is introduced. 
The solid line in figure 2 shows the fit achieved with the 2-component dark model
of Burkert and Silk adopting dark matter halo parameters which are in 
agreement with  CDM$\Lambda$ models  and using  a spherically symmetric
dark baryonic component
with a physically reasonable density distribution that decreases
monotonically with increasing radius. The required mass of the dark baryonic spheroid
is $1.5 \times 10^{9} M_{\odot}$ which is 4-5 times the mass of the visible
gaseous galactic disk and about 25\% the total mass of the non-baryonic dark component.
The apparent universality of rotation curves would then suggest that the relative
mix of the two  dark components should in turn be universal.
 The origin of such a dark baryonic
component which must have formed during an early dissipative 
condensation phase of baryons relative to the nondissipative, collisionless dark matter
component is not understood.
However this is not to say that it could not have occurred. Our understanding
of early star formation and the primordial initial mass function
is sufficiently primitive that this remains very much
an open possibility.

\end{document}